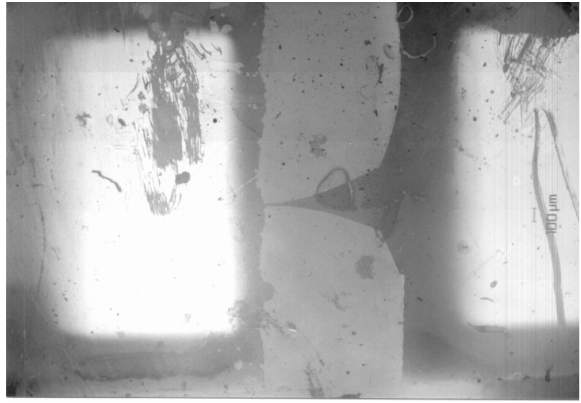

(a)

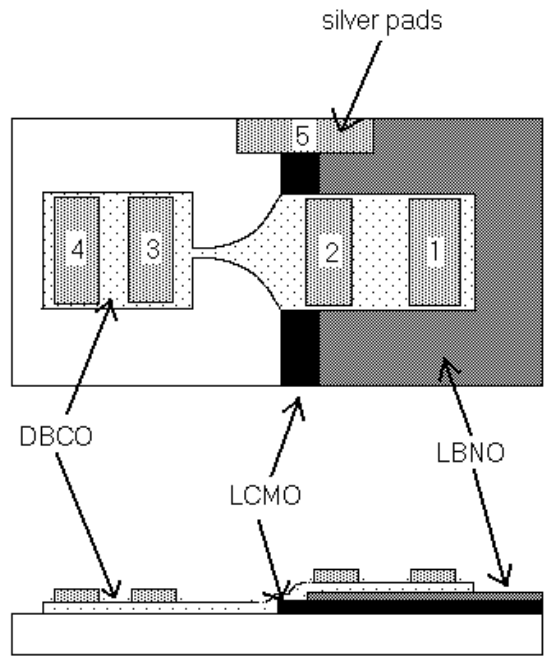

(b)

**Figure 1  P. Raychaudhuri *et al.***

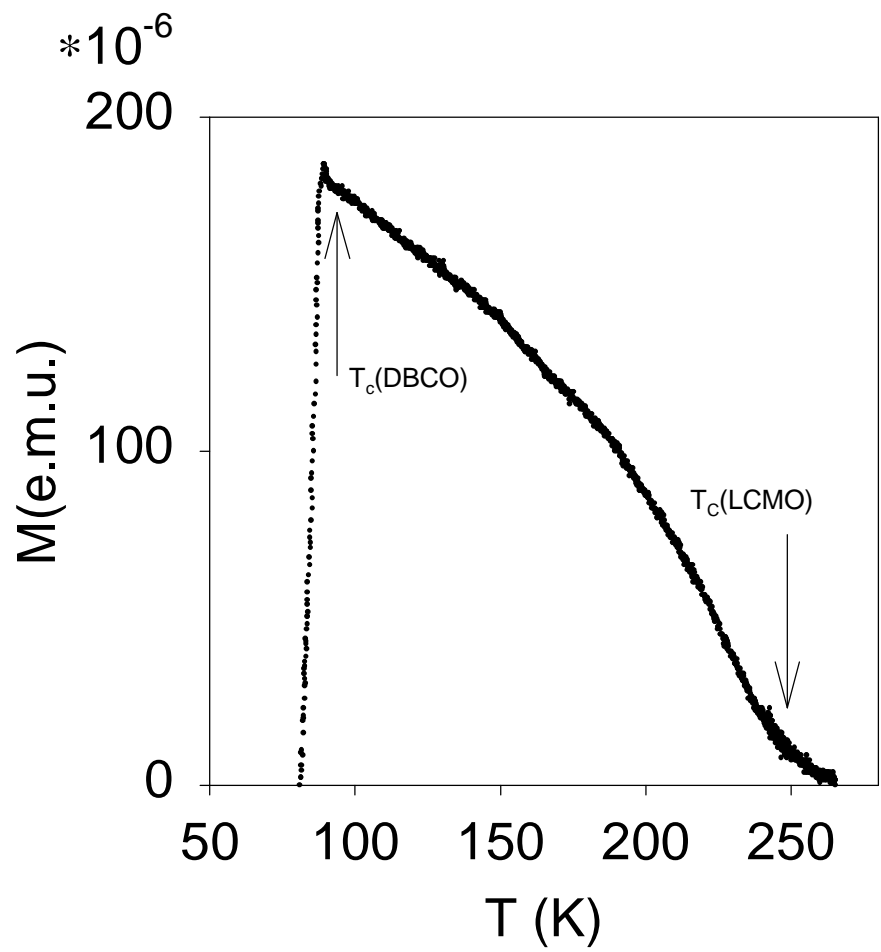

Figure 2  P. Raychaudhuri *et al.*

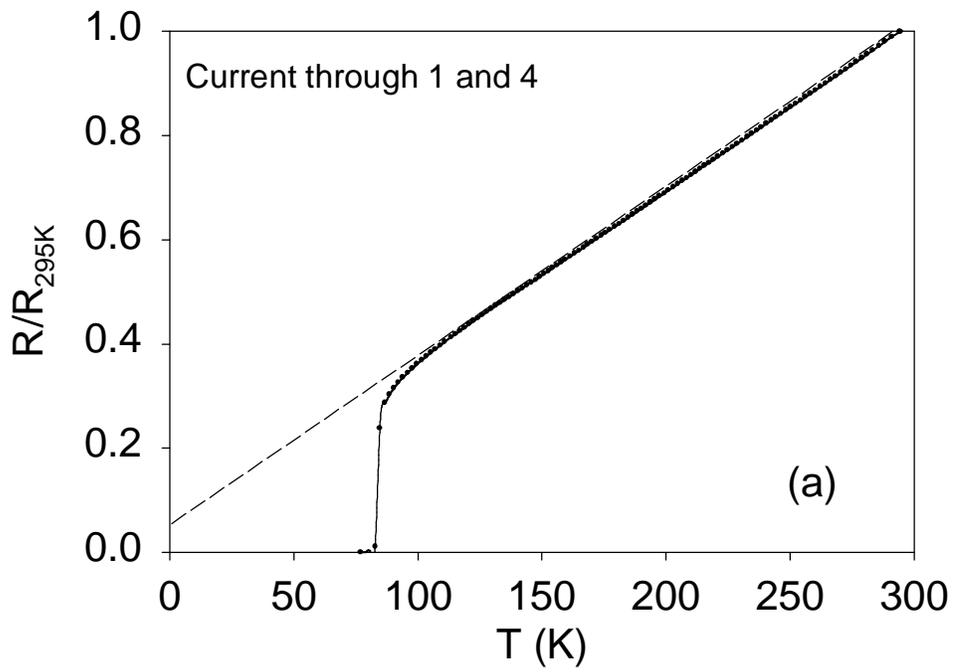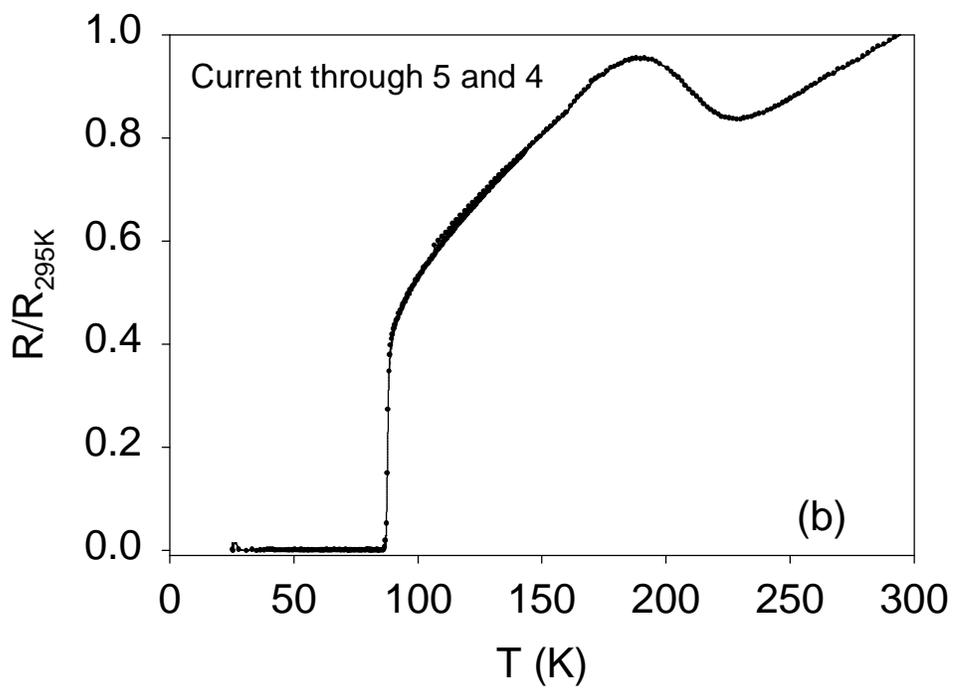

Figure 3   P. Raychaudhuri *et al.*

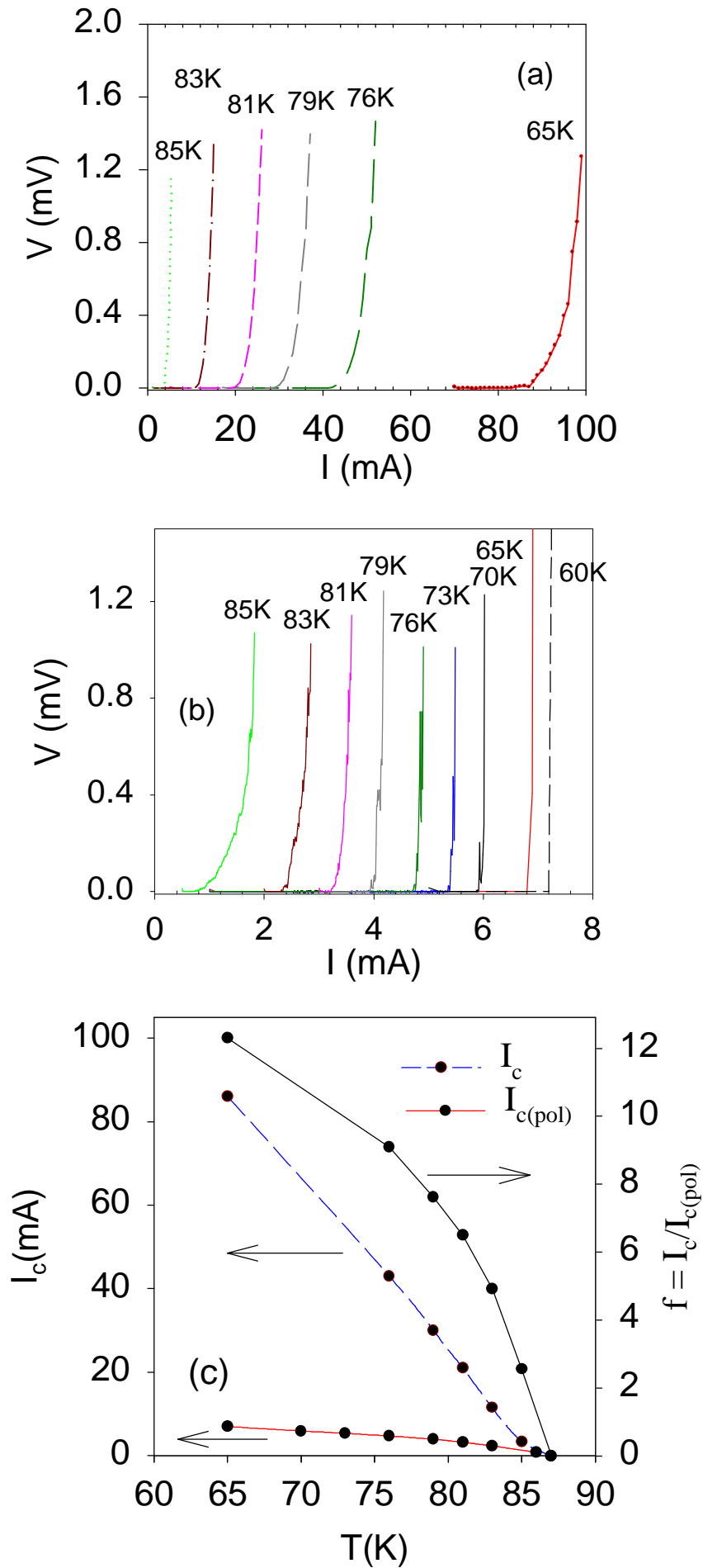

Figure 4 P. Raychaudhuri *et al*.

# Critical current of a superconductor measured via injection of spin polarized carriers

P. Raychaudhuri[a)], P. K. Mal[b)], S. Sarkar[c)], A. R. Bhangale and R. Pinto[d)]
*Department of Condensed Matter Physics and Materials Sciences,
Tata Institute of Fundamental Research,
Homi Bhabha Rd., Mumbai 400005,
India.*

*Abstract:* In this paper we report a direct evidence of the suppression of critical current due to pair-breaking in a superconducting micro-bridge when the measurement is carried out by injecting spin polarised carriers instead of normal electrons. A thin layer of $La_{0.7}Ca_{0.3}MnO_3$ was used as the source of spin polarised carriers. The micro-bridge was formed on the $DyBa_2Cu_3O_{7-\delta}$ thin film by photo-lithographic techniques. The design of our spin-injection device allowed us to inject spin-polarised carriers from the $La_{0.7}Ca_{0.3}MnO_3$ layer directly to the $DyBa_2Cu_3O_{7-\delta}$ micro-bridge (without any insulating buffer layer) making it possible to measure the critical current when polarised electrons alone are injected into the superconductor. Our results confirm the role of polarised carriers in breaking the Cooper pairs in the superconductor.

[a]e-mail: pratap@tifr.res.in
[b]e-mail: prolay@tifr.res.in
[c]e-mail: shampa@tifr.res.in
[d]e-mail: rpinto@tifr.res.in

The half-metallic nature[1,2] of hole doped rare-earth manganites of the form $R_{1-x}A_xMnO_3$ (R = rare-earth, A = bivalent cation) provides us with a reserve of spin polarised electrons whose charges as well as spins can be utilised by integrating them in unconventional devices. Towards this end, several prospective applications such as tunneling magnetoresistance devices with both positive and negative magnetoresistance[3,4] as well as "spin injection" devices using a high temperature superconducting layer on the top of a ferromagnetic layer have been proposed[5-9]. In a spin injection device, polarised carriers from a half metallic ferromagnetic layer are injected into the superconductor through a thin insulating layer. It has been demonstrated that this results in a suppression of critical current in the superconducting layer due to the breaking of the time-reversal symmetry of the Cooper pairs via the polarised electrons. This is analogous to the pair breaking effect caused by magnetic field in a superconductor.

Earlier experiments on spin injection in superconductors[7-9] were carried out by injecting the polarised spins from the ferromagnetic layer through a thin insulating barrier. The necessity of the insulating layer stemmed from the particular geometry of the devices used in those studies. However, the significant Joule heating due to the passage of the current through the insulating layers in those experiments restricted the magnitude of the injection current that could be passed from the ferromagnet to the superconductor. For the same reason the current versus voltage (I-V) characteristics by the passage of spin polarised quasiparticle current alone were not reported in those devices. Moreover, since in all these devices the superconducting layer was placed directly on the top of the ferromagnetic layer (and separated by an insulating barrier), vortex nucleation inside the superconductor due to the local field of the ferromagnetic layer could not be ruled out. That this is a possibility is suggested from the observed increase in critical current of the superconductor with increasing insulating layer thickness[9].

In this work, we report the I-V characteristic of the superconductor when spin polarised current alone is passed through a thin superconducting micro-bridge. The spin polarised current, in the present device is injected directly from the ferromagnet to the superconductor without any insulating barrier. The schematic diagram of the device used for our experiments is shown in figure 1a. The device was fabricated by first depositing a $La_{0.7}Ca_{0.3}MnO_3$ (LCMO) layer (1000Å) on half of the single crystalline $LaAlO_3$ substrate using pulsed laser deposition (PLD) while covering the remaining half by a metal (SS304) mask. Subsequently a thin insulating layer (300Å) of $La_2BaNbO_6$ (LBNO) was deposited on the LCMO layer using PLD by positioning the metal mask such that 0.5mm strip of LCMO near the middle of the substrate is not covered by the insulator. Subsequently, a superconducting layer of $DyBa_2Cu_3O_{7-\delta}$ (DBCO, 1000Å thickness) was deposited on the entire substrate. X-ray diffraction θ–2θ scans on the device confirmed that all three layers were oriented with c-axis perpendicular to the substrate. A micro-bridge (10μ wide) was patterned on the superconducting film (fig. 1b) using photolithography. The micro-bridge was positioned such that the narrowest region of the micro-bridge was directly on the $LaAlO_3$ substrate. This precaution was taken to remove possible proximity effect (which will be significant within a length of the order of the coherence length ξ) of the LCMO layer, on the critical current of the superconductor. Four silver pads were deposited by evaporation on the DBCO layer for attaching the leads (leads 1-4 in fig. 1a) to measure the critical current of the superconducting layer. A fifth silver pad (lead 5 in fig. 1a) was deposited on the bare LCMO layer and was used for the injection of the polarised spin. The large silver pads (more than 100 times in width compared to the micro-bridge) both on the superconductor and on the LCMO layer ensured that the heating due to contact resistance was kept to the minimum during the measurements. The normal critical current ($I_c$) of the superconductor was measured by passing the current between 1 and 4 and measuring the voltage between 2 and 3. The critical current with spin polarised current ($I_{c(pol)}$) was measured by injecting the current through 5 and 4 and measuring the voltage between 2 and 3. The insulating LBNO layer[10] ensured that the spin polarised quasi-particle were injected in the superconductor in the vicinity of the micro-bridge.

The ferromagnetic transition temperature ($T_C$) of the LCMO layer was measured by measuring the magnetisation as a function of temperature (M-T) using a vibrating sample magnetometer in a field of 4 kOe (Fig. 2). The $T_C$ determined from the maximum in the double derivative of the M-T curve was at 245K. Below 87K, there was a sharp drop in the magnetisation due to the superconducting transition ($T_c$) of the DBCO layer. The $T_c$ of the superconducting film was also confirmed from resistance versus temperature (R-T) measurement. Figure 3a shows the R-T curve of the superconducting layer measured by passing a current of 100μA through 1 and 4. The sharp transition (<2K) confirms the high quality of the superconducting film. No significant change in $T_c$ was observed when the measurement was done by passing the same current through 5 and 4 (Fig. 3b). In addition, the observation of a zero resistance rules out the possibility of any significant contact resistance between the LCMO and superconducting DBCO layer. This is expected because of the very good lattice match between the LCMO and DBCO films. The insensitivity of the measured $T_c$ on the injection current requires further consideration and will be discussed later.

Figure 4a and 4b show the I-V characteristic of the superconductor measured by passing normal current and spin polarised current respectively. The measurements were carried out by passing short pulses of current (up to 100mA), in order to avoid Joule heating at the current contacts. Two distinct features are observed when the I-V curves measured by passing polarised spins:

(i) The drastic suppression in critical current compared to the case when the measurements are carried out with normal current (passed between 1 and 4).

(ii) The sharp rise in the voltage when the current reaches the critical current.

Figure 4c shows the critical current as a function of temperature with normal ($I_c$) and spin polarised current ($I_{c(pol)}$). In the same figure we have plotted the ratio $I_c/I_{c(pol)}(=f)$ as a function of temperature, which is a measure of the pair breaking efficiency by the polarised spins. This quantity increases with decreasing temperature signifying that the pair breaking by spin polarised electrons becomes more efficient at low temperatures.

In a superconductor with no current the electrons are coupled in pairs via time reversal symmetry which requires that their wave-vectors are $\mathbf{k}\uparrow$ and $-\mathbf{k}\downarrow$ where $\uparrow$ and $\downarrow$ denote spin up and down respectively. When a normal current is passed through the superconductor the wave-vectors are modified as $-\mathbf{k}+\mathbf{q}\uparrow$ and $\mathbf{k}+\mathbf{q}\downarrow$ respectively, (where $\mathbf{q}$ is dependent on the drift velocity of the super-electrons). The critical momentum of the carrier up to which the superconductor can support a current is given by $\hbar/\sqrt{3}\xi(T)$[11], (where $\xi(T)$ is the Ginzburg-Landau coherence length at temperature T) above which the superconductivity gets destroyed due to the vanishing of the energy gap. In an actual measurement of critical current, this limit is rarely achieved in high temperature superconducting film due the large size of the micro-bridge compared to the coherence length and due to the presence of weak links arising from the grain boundaries in the sample[12]. This quantity varies with temperature as $(1-(T/T_c))^{3/2}$ (Strictly speaking this mean field relation holds only at temperature close to $T_c$.) In addition to this mechanism, in the presence of a spin polarised current, additional pair-breaking occurs by the destruction of the time reversal symmetry caused due to the imbalance caused by injecting polarised spins. This is somewhat similar to the pair breaking due to a localised magnetic impurity though significant difference exists. In the case of spin injection the polarised carriers will have a finite lifetime inside the sample after which the polarisation will vanish due to various scattering of the spin polarised quasi-particles inside the superconductors. The lifetime of the carriers is also expected to be a function of temperature. At temperatures close to $T_c$ the spin polarised quasi-particle lifetime is expected to be small due to strong scattering. As the temperature is decreased this lifetime increases causing the spin injection to be much more effective. From our data the increase in *f* with decreasing temperature clearly suggest that *pair breaking due to spin polarised electrons becomes the dominant mechanism as the temperature is decreased below $T_c$.* This can be easily understood considering the geometry of our device. The critical current is measured at the narrowest point of the micro-bridge, whereas the spin is injected from the other end. At temperatures close to $T_c$ the spin polarised quasi-particles will get destroyed before reaching the narrowest point. At low temperatures however more and more spin polarised quasi-particles will reach the narrowest point of the micro-bridge making the spin injection more and more effective. This picture is also supported from the observation that we do not observe any $T_c$ suppression when the R-T is measured by using a spin polarised current. Though the experimental data contains an uncertainty due to the existence of weak links between various grains in the superconductor, the qualitative behaviour of *f* is unlikely to get significantly modified by the presence of these junctions. A second source of uncertainty comes from the increase in spin-polarisation inside the ferromagnet as the temperature is lowered. However, all our measurements are carried out at temperatures less than <0.3$T_C$ of the ferromagnet. We have shown earlier that in half metallic ferromagnets, the polarisation $P=M_s(T)/M_s(T=0)=m$[13]. The change in m from T=0.3$T_C$ to T=0 is of the order of 10% and is unlikely to account for the large increase in *f* observed below the $T_c$ of the superconductor. Therefore, in a phenomenological way *f* reflects the temperature dependence of the spin polarised quasi-particle lifetime in the superconductor.

      Our experiment also removes many of the possible artefacts present in the earlier experiments. Firstly, since the critical current is measured at the narrowest region of the micro-bridge which is far from the ferromagnetic layer compared to the penetration length (λ) and coherence length (ξ) the proximity effect and self-injection of polarised carriers[9] is avoided in our geometry. Furthermore the absence of any insulating barrier between the LCMO and DBCO layer minimises the possibility of any Joule heating in the sample. The only uncertainty in our experiment stems from the fact that the spin injection and the

measured $I_c$ are not on the same point of the superconductor. Part of the polarised spins gets depolarised when they pass through the micro-bridge to the narrowest point due to their finite lifetime. This makes it difficult to quantify the exact amount of $I_c$ suppression in a superconductor *when a known amount of spin polarised electrons are present in a steady state*. This quantity can however be estimated by positioning the micro-bridge at various distances from the ferromagnetic layer. Such work is underway and will be reported later.

It is also interesting to note from the I-V curves that the current (I) increase much more sharply (fig. 4a-b) above the critical current when the measurement is carried out with polarised spins. At temperatures below 65K we were not able to record any data point between the 0V and the 5μV criterion used as a measure of $I_{c(pol)}$. Though we do not have at present a theoretical model to explain this observation, it can be inferred that the pair breaking due to polarised spins is much stronger above $I_{c(pol)}$ than the pair breaking by normal electrons above $I_c$.

In summary, we have fabricated a ferromagnet to superconductor spin-injection device where the polarised spins are directly injected into the superconductor. Our results strongly suggest that the lifetime of the polarised carriers inside the superconductor increases as we decrease the temperature below the superconducting transition temperature. We strongly believe that further experiments with spin injection would give very useful information regarding quasi particle scattering mechanisms in high temperature superconducting materials. In addition the construction of our device makes it possible to use this as a superconducting switch by controlling the current through leads 5 and 4. Detailed switching characteristics of such a device will be reported in a subsequent paper.

**Figure Captions:**

**Figure 1.** (a) Optical photograph of the superconducting micro-bridge fabricated through photolithography. (b) Schematic construction of the spin injection device.

**Figure 2.** Magnetisation versus temperature of the device measured on a vibrating sample magnetometer in a field of 4kOe. The arrows show the ferromagnetic and superconducting transition temperature of the LCMO and DBCO layers respectively.

**Figure 3.** Normalised resistance versus temperature of the superconducting micro-bridge measured by passing current through (a) 1 and 4; (b) 5 and 4.

**Figure 4.** I-V characteristics of the superconducting micro-bridge measured by passing current through (a) 1 and 4; (b) 5 and 4. (c) Variation of $I_c$, $I_{c(pol)}$ and $f= I_c/I_{c(pol)}$ as a function of temperature.